\documentclass[a4paper]{jpconf}
\usepackage{graphicx}
\begin{document}
\title{Results on the primary CR spectrum and composition reconstructed with the SPHERE-2 detector}

\author{R~A~Antonov$^{1}$, S~P~Beschapov$^{2}$, E~A~Bonvech$^{1}$, D~V~Chernov$^{1}$, T~A~Dzhatdoev$^{1,*}$, Mir~Finger$^{3}$, Mix~Finger$^{3}$, V~I~Galkin$^{4}$, N~V~Kabanova$^{2}$, A~S~Petkun$^{2}$, D~A~Podgrudkov$^{4}$, T~M~Roganova$^{1}$, S~B~Shaulov$^{2}$ and T~I~Sysoeva$^{2}$}

\address{$^{1}$ Lomonosov Moscow State University Skobeltsyn Institute of Nuclear Physics, Moscow, Russia}
\address{$^{2}$ P.N. Lebedev Physical Institute of the Russian Academy of Sciences, Moscow, Russia}
\address{$^{3}$ Charles University, Prague, Czech Republic}
\address{$^{4}$ Faculty of Physics, Lomonosov Moscow State University, Moscow, Russia}
\ead{$^{*}$ timur1606@gmail.com}

\begin{abstract}
First preliminary results of the balloon-borne experiment SPHERE-2 on the all-nuclei primary cosmic rays (PCR) spectrum and primary composition are presented. The primary spectrum in the energy range $10^{16}$--$5\cdot10^{17}$ eV was reconstructed using characteristics of Vavilov-Cherenkov radiation of extensive air showers (EAS), reflected from a snow surface. Several sources of systematic uncertainties of the spectrum were analysed. A method for separation of the primary nuclei' groups based on the lateral distribution function' (LDF) steepness parameter is presented. Preliminary estimate of the mean light nuclei' fraction $f_{30-150}$ at energies $3\cdot10^{16}$--$1.5\cdot10^{17}$ eV was performed and yielded $f_{30-150}$= (21$\pm$11) \%.
\end{abstract}

\section{Introduction}

Despite more than 50 years of the cosmic rays' (CR) energy spectrum measurements, a considerable uncertainty of the spectral shape in the ultrahigh energy domain ($E>{10}^{15}$ eV= 1 PeV) still exists (for recent results compilation see \cite{abb12}, fig. 15). The difference between spectra reconstructed by the various experimental groups is sometimes quite pronounced, and cannot be attributed to the statistical issues only. For example, the CR all-nuclei spectrum measured by the GAMMA group \cite{gar08} exhibits statistically significant ($> 4\sigma$) sharp peak at 8$\cdot$10$^{16}$ eV. A similar feature exists in the spectrum based on the 2009--2010 run of the Tunka-133 experiment \cite{ant11a}--\cite{ant12a}, but for the 2010--2011 run of the same experiment it was not confirmed \cite{ant12a}. Also, the peak is not seen by other CR experiments such as Akeno \cite{nag92}, Tibet-III \cite{ame08} and KASCADE-Grande \cite{ape11}--\cite{ape12}.

The situation with the primary CR composition at superhigh energies is even more enigmatic, as the consensus between the different measurements, such as \cite{ant05}--\cite{tok08}, is virtually absent (see fig. 1 of \cite{tsu08}). More recent review \cite{kam12} claims that the results on the mean logarithmic mass number, $<ln A>$, are still consistent with a trend of growing at energies $E>$3-4 PeV, where the primary CR spectrum exhibits a pronounced break, the so-called knee \cite{kul58}. But anyway, the scatter of $<ln A>$ values measured by the different experiments is high, and more detailed information than just $<ln A>$ is needed for the detailed comparison with the CR acceleration and propagation models.

These examples highlight the importance of the all-nuclei CR spectrum measurement and composition studies in various experiments and by the diverse methods. In present work we present the first results of the SPHERE-2 experiment. The SPHERE-2 experiment is a Cherenkov telescope situated above the snow surface of Lake Baikal and registers reflected Vavilov-Cherenkov radiation ("Cherenkov light"). The basic idea of such an experiment was proposed in \cite{chu72}.

\section{The data}

The SPHERE-2 experiment uses a mosaic of 109 PMTs for recording the light signal reflected from the spherical mirror. Observation level is typically 400--900 m above the snow surface. In the present work we use the data from the 2011 and 2012 experimental runs. Time sampling step is 25 ns for the 2011 run and 12.5 ns for the 2012 run. For more detailed description of the SPHERE-2 experiment see \cite{ant11b}--\cite{ant12b}.

In the four flights of the 2011 run 230 events consistent with EAS Cherenkov light reflected from the snow surface were found. For the 2012 run we use the data taken in the first four flights, and the number of events classified as showers is 301. Relatively small number of the showers compared to the ground-based arrays is due to a high threshold of the EAS observation, as, in average, only 1 out of each $\sim$10$^6$ Cherenkov photons reflected from the snow surface reaches the detector's mirror.

\section{A method for the primary spectrum and composition reconstruction}

In order to reconstruct the primary CR spectrum and composition we applied an approach based on detector response simulation by means of the full Monte Carlo method. The procedure of the all-nuclei spectrum measurement consists of the several stages.

1. As the first step, the full Monte Carlo simulation of the lateral distribution function (LDF) of the EAS Cherenkov light, as well as the time structure of the shower was performed using the standard code CORSIKA6.500 \cite{hec98} (for more details see \cite{ano09}--\cite{dzh11}). All results presented here are based on the QGSJET-I high energy hadronic model \cite{kal97} and the GHEISHA low energy hadronic model \cite{fes85}.

2. Then, the geometrical and optical effects smearing the observed signal were accounted for. For the 2011 run reconstruction we use the simplified detector response model developed in \cite{ant09}--\cite{dzh11} and written in C++. In the case of the 2012 run the more advanced model is used, which was implemented using the standard code Geant4 \cite{ago03}.

3. The simulation of the instrumental acceptance was performed. The result is most sensitive to the observation altitude's change, as well as to the variations of the thresholds of the detector's channels.

4. For the sample of the EASs registered in the experiment the primary energy was estimated. We use the method of \cite{ded04}, where the energy estimation was performed together with the axis position reconstruction by normalising of the experimental LDFs to the model LDFs with known energy.

5. The all-nuclei CR spectrum was reconstructed using the sample of the showers with estimated energy and results of the acceptance calculation.

The primary composition study in the present work is based on the LDF steepness parameter, defined as the ratio of the number of Cherenkov photons in the circle with the radius of 67 m to the same number in the ring with the radii of 67 m and 134 m \cite{ant09}. The parameter allows to select some fraction of the light nuclei \cite{ano09}--\cite{ant09}. In present work protons stand for light nuclei, and Iron for heavy ones.

\section{Preliminary results}

The all-nuclei CR spectrum reconstructed using 2011--2012 runs data of the SPHERE-2 experiment is shown in fig. 1 (red stars). Statistical (red bars) and systematic (red dashed lines) uncertainties of the spectrum are shown. The main sources of systematic uncertainty of the spectrum are as follows: a) bin-to-bin migration due to the energy estimation error (dominant factor at E$>$20 PeV); b) acceptance estimation inaccuracy (not more than 5 \%); c) the spectal shape vs. the primary composition dependence (dominates at E$<$20 PeV). With respect to the direct predecessor of the present work \cite{ant97}, we achieved a considerable progress of the reconstruction methodology. Also the results of some other experiments are shown in fig. 1a for comparison: Akeno \cite{nag92} (green circles), KASCADE-Grande \cite{ape11} (black triangles), and Tunka-133 \cite{ber11} (blue squares). Statistical errors of the Akeno spectrum are comparable with the circle's diameter; systematic uncertainties of the KASCADE-Grande result are shown by black dashed lines.

Estimated fraction of light nuclei $f$ vs. $lg(E)$ is shown in fig. 2 (red circles) with statistical uncertainties of the $f$ values (red bars). Several  independent reconstructions of the experimental showers' LDFs, and then the $f$ vs. $lg(E)$ dependence were performed, and induced systematic uncertainty of the $f$ value was estimated (black lines). In the first two bins the value of $f$ is strongly modified by the threshold effects; an additional uncertainty assotiated with them is shown in the picture by blue arrows. The data presented in fig. 2 makes it possible to estimate the mean (not intensity-weighted!) fraction of light nuclei in the energy region 30--150 PeV $f_{30-150}= 0.21 \pm 0.11$. The corresponding value of $<ln A>= 3.20 \pm 0.45$ (comparable with values allowed by \cite{kam12}; see fig. 17).

All presented results are subject to additional recheck and develompent of the methodology, and the detailed study of the systematic uncertainties that might change the results presented at figs. 1--2 is underway.

\begin{figure}[b] 
\vspace{-1.0pc}
\begin{minipage}{18pc}
\includegraphics[width=20pc]{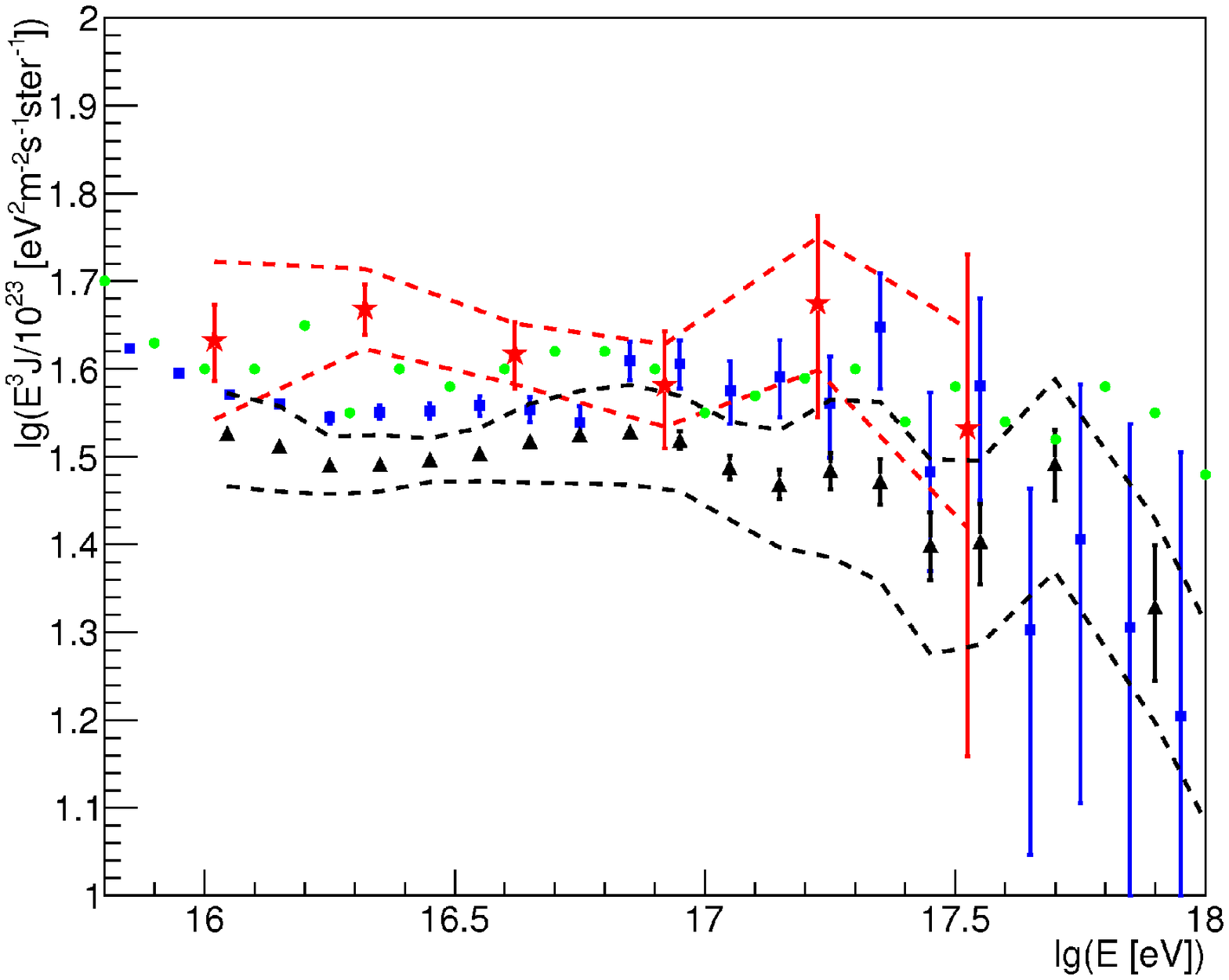}
\caption{\label{label1} (color online). The all-nuclei CR spectrum measured in the SPHERE-2 experiment (red stars; preliminary result) together with the statistical and the systematic uncertainties, as well as results of the Akeno \cite{nag92} (green circles), KASCADE-Grande \cite{ape11} (black triangles) and Tunka-133 \cite{ber11} experiments (blue squares).}
\end{minipage}\hspace{2pc}%
\begin{minipage}{18pc}
\includegraphics[width=20pc]{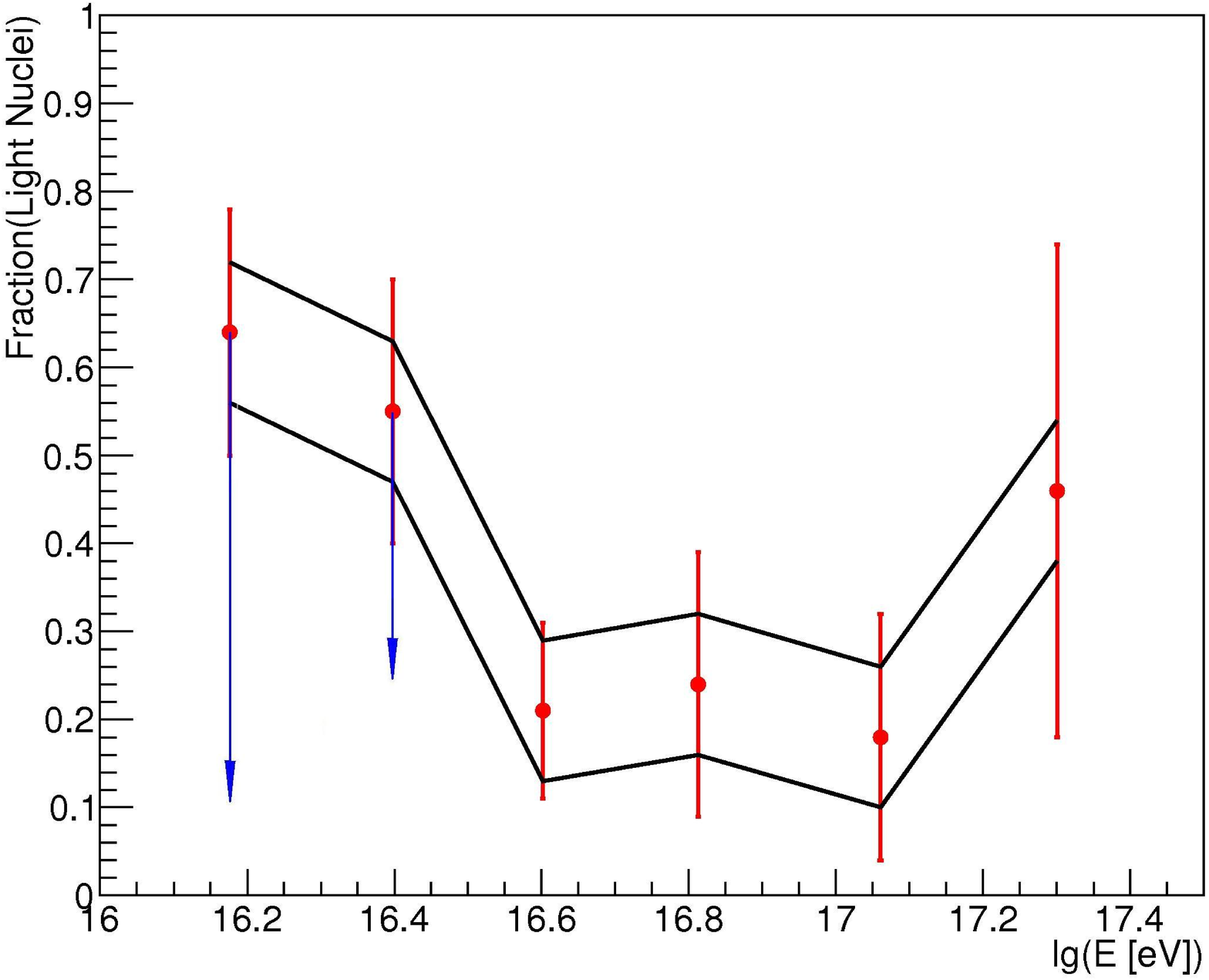}
\caption{\label{label2} (color online). The reconstructed fraction of light nuclei $f$ vs. $lg(E)$, as well as the statistical (red bars) and the systematic (black lines) uncertainties of $f$. Also the additional error of the light-nuclei fraction arising from the threshold effects (blue arrows) is shown.}
\end{minipage}
\vspace{-1.0pc}
\end{figure}

\section{Conclusions}

In the present work we performed the first detailed reconstruction of the CR all-nuclei spectrum using Cherenkov light reflected from the snow surface, as well as the very first study of the primary composition by means of this method. Considerable effort was devoted to establishing the database of the full Monte Carlo model showers in order to correctly simulate their fluctuations. A dependence of the spectral shape vs. primary composition in the threshold region was accounted for. It was found that the overall fraction of light nuclei in the energy region 30--150 PeV $f_{30-150}= 0.21 \pm 0.11$. All above-mentioned results are preliminary and are subject to the additional recheck.

To conclude, let us note that the developed method of the all-nuclei spectrum reconstruction and composition studies might be useful for analysing of the next generation CR experiments capable of observation of Cherenkov light reflected from the natural surfaces, particularly, for orbital experiments such as TUS \cite{shu11} and JEM-EUSO \cite{tak09}.

\section{Acknowledgements}
The authors acknowledge the Russian Foundation for Basic Research (grant 11-02-01475-a, 12-02-10015-k, LSS-871.2012.2) and the Program of basic researches of the Presidium of the Russian Academy of Sciences "Fundamental properties of matter and astrophysics" for the support of the research. Authors are grateful to the technical collaborators of the SPHERE-2 experiment. Calculations of the instrumental acceptance were performed using the SINP MSU high performance computer cluster. Work of T. Dzhatdoev was supported by the MSU Research Council grant for young researchers.

\smallskip
\end{document}